\makeatletter
\@ifundefined{@parse@version@dash}{%
\def\@parse@version#1{\@parse@version@0#1}
\def\@parse@version@#1/#2/#3#4#5\@nil{%
\@parse@version@dash#1-#2-#3#4\@nil}
\def\@parse@version@dash#1-#2-#3#4#5\@nil{%
  \if\relax#2\relax\else#1\fi#2#3#4 }
}{}
\makeatother
\documentclass[reprint,showkeys,longbibliography,nofootinbib,superscriptaddress,aps,amsmath,amssymb,notitlepage]{revtex4-2}
\usepackage {xcolor}

\usepackage{graphicx}% Include figure files
\usepackage{dcolumn}% Align table columns on decimal point
\usepackage{bm}% bold math
\usepackage{multirow}
\allowdisplaybreaks
\usepackage{verbatim}
\usepackage{xspace} 
\newcommand{\sige}{\mbox{$\bar\sigma_e$}}
\newcommand{\mass}{\mbox{$m_\chi$}}
\newcommand{\crystal}{\mbox{$f_\textnormal{c}(q,E_e)$}}

\newcommand{\geant}{\textsc{Geant4}\xspace}

\begin{document}
%\preprint{APS/123-QED}

\title{First Constraints from DAMIC-M on Sub-GeV Dark-Matter Particles Interacting with Electrons}
% !TEX root = main.tex
%\email[Corresponding author: ]{spokesperson@email}

% alphabetical

\author{I.\,Arnquist}
\affiliation{Pacific Northwest National Laboratory (PNNL), Richland, WA, United States} 

\author{N.\,Avalos}
\affiliation{Centro At\'{o}mico Bariloche and Instituto Balseiro, Comisi\'{o}n Nacional de Energ\'{i}a At\'{o}mica (CNEA), Consejo Nacional de Investigaciones Cient\'{i}ficas y T\'{e}cnicas (CONICET), Universidad Nacional de Cuyo (UNCUYO), San Carlos de Bariloche, Argentina}

\author{D.\,Baxter}\email[Now at Fermi National Accelerator Laboratory, Batavia, IL, United States]{}
\affiliation{Kavli Institute for Cosmological Physics and The Enrico Fermi Institute, The University of Chicago, Chicago, IL, United States}

\author{X.\,Bertou}
\affiliation{Centro At\'{o}mico Bariloche and Instituto Balseiro, Comisi\'{o}n Nacional de Energ\'{i}a At\'{o}mica (CNEA), Consejo Nacional de Investigaciones Cient\'{i}ficas y T\'{e}cnicas (CONICET), Universidad Nacional de Cuyo (UNCUYO), San Carlos de Bariloche, Argentina}

\author{N.\,Castell\'{o}-Mor}
\affiliation{Instituto de F\'{i}sica de Cantabria (IFCA), CSIC - Universidad de Cantabria, Santander, Spain}

\author{A.E.\,Chavarria}
\affiliation{Center for Experimental Nuclear Physics and Astrophysics, University of Washington, Seattle, WA, United States}

%\author{N.J.\,Corso}
\author{J.\,Cuevas-Zepeda}
\affiliation{Kavli Institute for Cosmological Physics and The Enrico Fermi Institute, The University of Chicago, Chicago, IL, United States}

\author{J.\,Cortabitarte Guti\'{e}rrez}
\affiliation{Instituto de F\'{i}sica de Cantabria (IFCA), CSIC - Universidad de Cantabria, Santander, Spain}

\author{J.\,Duarte-Campderros}
\affiliation{Instituto de F\'{i}sica de Cantabria (IFCA), CSIC - Universidad de Cantabria, Santander, Spain}

\author{A.\,Dastgheibi-Fard}
\affiliation{LPSC LSM, CNRS/IN2P3, Universit\'{e} Grenoble-Alpes, Grenoble, France}

\author{O.\,Deligny}
\affiliation{CNRS/IN2P3, IJCLab, Universit\'{e} Paris-Saclay, Orsay, France}

\author{C.\,De Dominicis}
\affiliation{Laboratoire de physique nucl\'{e}aire et des hautes \'{e}nergies (LPNHE), Sorbonne Universit\'{e}, Universit\'{e} Paris Cit\'{e}, CNRS/IN2P3, Paris, France}
\affiliation{SUBATECH, Nantes Universit\'{e}, IMT Atlantique, CNRS-IN2P3, Nantes, France}

\author{E.\,Estrada}
\affiliation{Centro At\'{o}mico Bariloche and Instituto Balseiro, Comisi\'{o}n Nacional de Energ\'{i}a At\'{o}mica (CNEA), Consejo Nacional de Investigaciones Cient\'{i}ficas y T\'{e}cnicas (CONICET), Universidad Nacional de Cuyo (UNCUYO), San Carlos de Bariloche, Argentina}

\author{N.\,Gadola}
\affiliation{Universit\"{a}t Z\"{u}rich Physik Institut, Z\"{u}rich, Switzerland}

\author{R.\,Ga\"{i}or}
\affiliation{Laboratoire de physique nucl\'{e}aire et des hautes \'{e}nergies (LPNHE), Sorbonne Universit\'{e}, Universit\'{e} Paris Cit\'{e}, CNRS/IN2P3, Paris, France}

\author{T.\,Hossbach}
\affiliation{Pacific Northwest National Laboratory (PNNL), Richland, WA, United States} 

\author{L.\,Iddir}
\affiliation{Laboratoire de physique nucl\'{e}aire et des hautes \'{e}nergies (LPNHE), Sorbonne Universit\'{e}, Universit\'{e} Paris Cit\'{e}, CNRS/IN2P3, Paris, France}

\author{L.\,Khalil}
\affiliation{Laboratoire de physique nucl\'{e}aire et des hautes \'{e}nergies (LPNHE), Sorbonne Universit\'{e}, Universit\'{e} Paris Cit\'{e}, CNRS/IN2P3, Paris, France}

\author{B.\,Kilminster}
\affiliation{Universit\"{a}t Z\"{u}rich Physik Institut, Z\"{u}rich, Switzerland}

\author{A.\,Lantero-Barreda}
\affiliation{Instituto de F\'{i}sica de Cantabria (IFCA), CSIC - Universidad de Cantabria, Santander, Spain}

\author{I.\,Lawson}
\affiliation{SNOLAB, Lively, ON, Canada }

\author{S.\,Lee}
\affiliation{Universit\"{a}t Z\"{u}rich Physik Institut, Z\"{u}rich, Switzerland}

\author{A.\,Letessier-Selvon}
\affiliation{Laboratoire de physique nucl\'{e}aire et des hautes \'{e}nergies (LPNHE), Sorbonne Universit\'{e}, Universit\'{e} Paris Cit\'{e}, CNRS/IN2P3, Paris, France}

\author{P.\,Loaiza}
\affiliation{CNRS/IN2P3, IJCLab, Universit\'{e} Paris-Saclay, Orsay, France}

\author{A.\,Lopez-Virto}
\affiliation{Instituto de F\'{i}sica de Cantabria (IFCA), CSIC - Universidad de Cantabria, Santander, Spain}

\author{A.\,Matalon}
\affiliation{Kavli Institute for Cosmological Physics and The Enrico Fermi Institute, The University of Chicago, Chicago, IL, United States}
\affiliation{Laboratoire de physique nucl\'{e}aire et des hautes \'{e}nergies (LPNHE), Sorbonne Universit\'{e}, Universit\'{e} Paris Cit\'{e}, CNRS/IN2P3, Paris, France}

\author{S.\,Munagavalasa}
\affiliation{Kavli Institute for Cosmological Physics and The Enrico Fermi Institute, The University of Chicago, Chicago, IL, United States}

\author{K.J.\,McGuire}
\affiliation{Center for Experimental Nuclear Physics and Astrophysics, University of Washington, Seattle, WA, United States}

\author{P.\,Mitra}
\affiliation{Center for Experimental Nuclear Physics and Astrophysics, University of Washington, Seattle, WA, United States}

\author{D.\,Norcini}
\affiliation{Kavli Institute for Cosmological Physics and The Enrico Fermi Institute, The University of Chicago, Chicago, IL, United States}

\author{G.\,Papadopoulos}
\affiliation{Laboratoire de physique nucl\'{e}aire et des hautes \'{e}nergies (LPNHE), Sorbonne Universit\'{e}, Universit\'{e} Paris Cit\'{e}, CNRS/IN2P3, Paris, France}

\author{S.\,Paul}
\affiliation{Kavli Institute for Cosmological Physics and The Enrico Fermi Institute, The University of Chicago, Chicago, IL, United States}

\author{A.\,Piers}
\affiliation{Center for Experimental Nuclear Physics and Astrophysics, University of Washington, Seattle, WA, United States}

\author{P.\,Privitera}
\affiliation{Kavli Institute for Cosmological Physics and The Enrico Fermi Institute, The University of Chicago, Chicago, IL, United States}

\affiliation{Laboratoire de physique nucl\'{e}aire et des hautes \'{e}nergies (LPNHE), Sorbonne Universit\'{e}, Universit\'{e} Paris Cit\'{e}, CNRS/IN2P3, Paris, France}

\author{K.\,Ramanathan}\email[Now at the California Institute of Technology, Pasadena, CA, United States]{}
\affiliation{Kavli Institute for Cosmological Physics and The Enrico Fermi Institute, The University of Chicago, Chicago, IL, United States}

\author{P.\,Robmann}
\affiliation{Universit\"{a}t Z\"{u}rich Physik Institut, Z\"{u}rich, Switzerland}

\author{M.\,Settimo}
\affiliation{SUBATECH, Nantes Universit\'{e}, IMT Atlantique, CNRS-IN2P3, Nantes, France}

\author{R.\,Smida}
\affiliation{Kavli Institute for Cosmological Physics and The Enrico Fermi Institute, The University of Chicago, Chicago, IL, United States}

\author{R.\,Thomas}
\affiliation{Kavli Institute for Cosmological Physics and The Enrico Fermi Institute, The University of Chicago, Chicago, IL, United States}

\author{M.\,Traina}
\affiliation{Center for Experimental Nuclear Physics and Astrophysics, University of Washington, Seattle, WA, United States}
\affiliation{Laboratoire de physique nucl\'{e}aire et des hautes \'{e}nergies (LPNHE), Sorbonne Universit\'{e}, Universit\'{e} Paris Cit\'{e}, CNRS/IN2P3, Paris, France}

\author{I.\,Vila}
\affiliation{Instituto de F\'{i}sica de Cantabria (IFCA), CSIC - Universidad de Cantabria, Santander, Spain}

\author{R.\,Vilar}
\affiliation{Instituto de F\'{i}sica de Cantabria (IFCA), CSIC - Universidad de Cantabria, Santander, Spain}

\author{G.\,Warot}
\affiliation{LPSC LSM, CNRS/IN2P3, Universit\'{e} Grenoble-Alpes, Grenoble, France}

\author{R.\,Yajur}
\affiliation{Kavli Institute for Cosmological Physics and The Enrico Fermi Institute, The University of Chicago, Chicago, IL, United States}

\author{J-P.\,Zopounidis}
\affiliation{Laboratoire de physique nucl\'{e}aire et des hautes \'{e}nergies (LPNHE), Sorbonne Universit\'{e}, Universit\'{e} Paris Cit\'{e}, CNRS/IN2P3, Paris, France}

\collaboration{DAMIC-M Collaboration}

\date{\today}% 

\begin{abstract}
We report constraints on sub-GeV dark matter particles interacting with electrons from the first underground operation of DAMIC-M detectors. The search is performed with an integrated exposure of 85.23\,g\,days, and exploits the sub-electron charge resolution and low level of dark current of DAMIC-M Charge-Coupled Devices (CCDs).  
Dark-matter-induced ionization signals above the detector dark current are searched for in CCD pixels with charge up to 7\,e$^-$. 
With this data set we place limits on dark matter particles of mass between 0.53 and 1000\,MeV$/c^2$, excluding unexplored regions of parameter space in the mass ranges [1.6,1000]\,MeV$/c^2$ and [1.5,15.1]\,MeV$/c^2$ for ultra-light and heavy mediator interactions, respectively.
\end{abstract}

\maketitle

There is overwhelming evidence indicating that our universe is dominated by non-luminous, non-baryonic dark matter (DM)\,\cite{Zwicky:1933,Rubin:1970,Ostriker:1974}. The contemporary Standard Model of Cosmology, $\Lambda$CDM, is consistent with the observed cosmic background radiation features\,\cite{Planck2020} and large-scale distribution of galaxies\,\cite{eboss} when parameterized with a cold-DM particle density; precise measurements of Milky Way stars dynamics\,\cite{Salomon_2020} determine the local DM density. Many hypothetical particle candidates with the required properties\,\cite{Bertone_2005} have been proposed, however DM has yet to be directly detected. Motivated by a weak-scale annihilation cross section to explain today's measured abundance, searches for Weakly Interacting Massive Particles (WIMPs) with masses larger than the proton's ($\approx 1$~GeV$/c^2$) have been leading the experimental landscape. However, with null results from these multi-tonne detectors\,\cite{Akerib_2017,Aprile_2018,Meng_2021}, low-threshold experiments have been developed to search for light (sub-GeV) DM, including light WIMPs and hidden-sector particles\,\cite{USCosmicVisions2017}. Such detectors are designed to be sensitive to both sub-keV nuclear recoils and eV-scale electronic recoils induced by DM scattering. The latter scenario gives access to possible hidden-sector DM candidates that interact via a new gauge boson which is feebly mixed with the photon~\cite{Holdom1986,Okun1982}. Such a mixing provides a mechanism for DM-e$^-$ scattering to occur. 

The DAMIC-M (Dark Matter in CCDs at Modane) experiment\,\cite{damic_idm2022} searches for sub-GeV DM using skipper Charge-Coupled Devices (CCDs) under the French Alps at the Laboratoire Souterrain de Modane (LSM). DM-induced ionization events in the thick silicon bulk can be detected with sub-electron resolution through non-destructive, repeated pixel readout\,\cite{skipper,Chandler1990zz,Tiffenberg:2017aac,Compton-DAMICM}. Combined with an extremely low dark current\,\cite{Damic_dc_2019,sensei2020}, sensitivity to single-electron measurements allows DAMIC-M to achieve an energy threshold of a few\,eV. The completed experiment will feature $\approx$700\,g of target mass with an expected total background of a fraction of a dru\,\footnote{1\,dru = 1\,event/kg/keV/day.}. A prototype detector, the Low Background Chamber (LBC), is currently operating at LSM. The LBC aims to demonstrate the performance of the CCDs, background control strategy, and the sensitivity to light dark matter.

In this letter, we present the first search for sub-GeV DM with the DAMIC-M LBC. With an integrated exposure of 85.23\,g days, we set world-leading limits on dark matter-electron scattering interactions via heavy and ultralight mediators. The DM interaction model, data-taking conditions, and analysis strategy are detailed in the following. 

Theoretical expectations for hidden-sector DM interactions in crystalline silicon are derived in Refs.\,\cite{QEDark,Lee_2015,Griffin_2021,Knapen_2022}. The differential event rate from DM-e$^-$ interactions in the detector for a DM mass \mass\ with recoil energy $E_e$ is parameterized as~\cite{QEDark} 

\begin{equation}
    \label{eq:semicondrate}
    \frac{dR}{dE_e} \propto \bar\sigma_e \int \frac{dq}{q^2}\eta(m_\chi,q,E_e)|F_\text{DM}(q)|^2|f_c(q,E_e)|^2, 
\end{equation}

\noindent where \sige\ is a model-independent reference cross section for DM-e$^-$ elastic scattering, $q$ is the transferred momentum, $\eta$ includes properties of the incident flux of galactic DM particles, $F_{DM}$ is the DM form factor, and \crystal\ quantifies the atomic transitions of bound-state electrons\,\cite{QEDark}.
The DM form factor $F_{DM}=(\alpha m_e/q)^n$, where $\alpha$ is the fine-structure constant and $m_e$ the electron mass, describes the momentum-transfer dependence of the interaction, with $n=0$ for a point-like interaction with heavy mediators (mass $\gg\alpha m_e$) or a magnetic dipole coupling, $n=1$ for an electric dipole coupling, and $n=2$ for massless or ultra-light mediators (mass $\ll\alpha m_e$). The crystal form factor $f_c$, which includes the material properties of the silicon target, is calculated numerically with a DFT (density functional theory) approach (see Refs.\,\cite{QEDark,quantumespresso}).

\begin{figure}[t!]
    \centering
    \includegraphics[width=\linewidth]{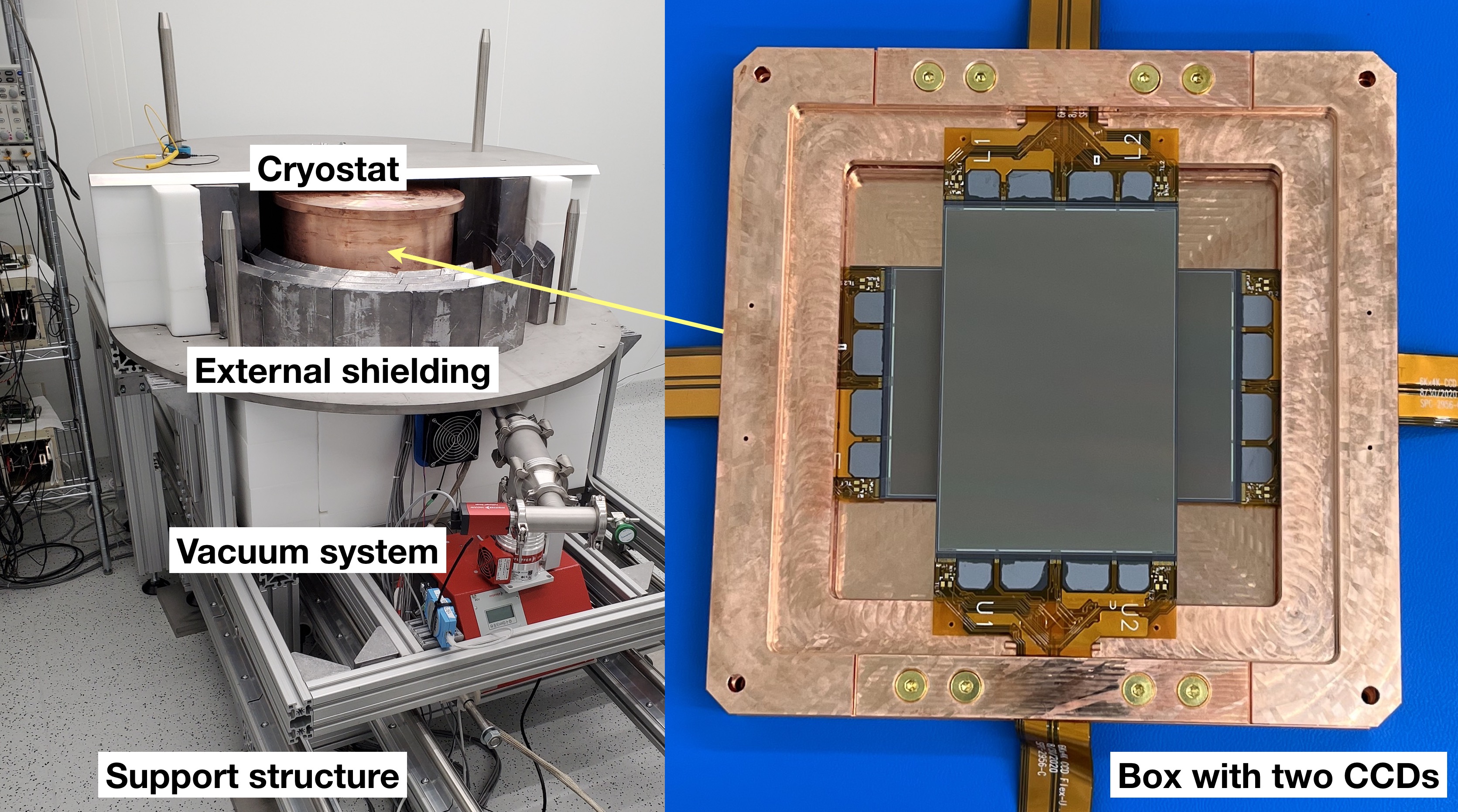}
    \caption{The DAMIC-M Low Background Chamber installed underground at LSM: the two skipper CCDs are mounted in a high-purity copper box (right); the box is placed inside the copper cryostat, visible here (left) during assembly of the external lead and polyethylene shielding.}
    \label{fig:LBC-CCD}
\end{figure}

Data are collected with two large-area, thick CCDs featuring 6144\,$\times$\,4128 pixels, as shown in Fig.\,\ref{fig:LBC-CCD}(right). Each pixel is a 15$\times$15\,$\mathrm{\mu m^{2}}$ square with a thickness of 670\,$\mathrm{\mu m}$, such that the total target mass per CCD is $\approx$9\,g. The CCD has a three-phase polysilicon gate structure with a buried p-channel, where charge carriers collected from fully depleted high-resistivity ($>$10\,k$\Omega$\,cm) n-type silicon bulk are clocked toward a readout amplifier\,\cite{Holland:2002zz,Holland:2003zz,Holland:2009zz}. Flex cables wire-bonded to the CCD provide the required voltage biases and clocks.

The two CCDs are mounted in a high-purity, oxygen-free, high-conductivity copper box, which also acts as a shield to infrared radiation. 
To minimize leakage current, the CCDs are operated at low temperature ($\approx 130$\,K) under vacuum (pressure $\sim$5$\times$10$^{-6}$\,mbar) inside the LBC cryostat, as in Fig.\,\ref{fig:LBC-CCD}(left). 
The CCD box is surrounded by at least 7.5\,cm of very low-background lead ($\le$ 7 mBq/kg $^{210}$Pb), with the innermost 2\,cm of ancient origin, to mitigate gamma radiation from components located in the cryostat: cables, electronics, fasteners, and cryocooler.
In addition, 15\,cm of low-background lead (54 Bq/kg $^{210}$Pb) and 20\,cm of high-density polyethylene surround the cryostat to attenuate high-energy $\gamma$-rays and neutrons, as shown in Fig.\,\ref{fig:LBC-CCD}(left). All parts of the detector are appropriately cleaned to remove any surface contamination~\cite{HOPPE2007486,ABGRALL201622}. A full simulation of the apparatus with \geant\,\cite{geant4} gives an expected total background of $\sim$10\,dru for this initial LBC installation. The simulation includes realistic amounts of radioactive contaminants as determined by radioassay measurements and bookkeeping of cosmogenic activation time of materials (see Ref.\,\cite{PhysRevD.105.062003} for similar methods). This level of background, similar to that achieved by DAMIC at SNOLAB\,\cite{PhysRevD.105.062003}, was confirmed during the LBC commissioning and has negligible impact on the analysis presented in this letter. Voltage biases and clocks to operate the devices are provided by a commercial CCD controller from Astronomical Research Cameras, Inc. placed outside the external shielding.

A DM-e$^-$ interaction in the bulk silicon of the CCD will generate charge carriers in numbers proportional to the energy deposited. The voltage bias applied for full depletion of the substrate (70~V) drifts the charge along the z-direction toward the x-y plane of the CCD pixel array. Thermal diffusion in the transverse direction results in a spatial variance of the charge collected at the pixel array, $\mathrm{\sigma^{2}_{xy}}$, proportional to the transit time\,\cite{PhysRevD.94.082006}. To read out the pixel, charge is moved by voltage clocks, first vertically row-by-row toward the serial register of the CCD, and then horizontally, to the two charge-to-voltage amplifiers (referred to as U and L) at each end of the serial register. DAMIC-M CCDs feature skipper amplifiers\,\cite{skipper,Chandler1990zz,Tiffenberg:2017aac}, which can be configured to make multiple, non-destructive charge measurements (NDCMs). The charge resolution improves as $1/\sqrt{\rm{N_{skip}}}$, where $\rm{N_{skip}}$ is the number of NDCMs, when averaging all the measurements,  reaching a sub-electron level for sufficiently large $\rm{N_{skip}}$. 
Details on the performance of DAMIC-M skipper CCDs can be found in Ref.\,\cite{Compton-DAMICM}.

After commissioning of the CCDs -- which includes optimizing the operating parameters for charge transfer efficiency, resolution, and dark current -- two data sets with similar exposures are collected between May and July 2022. An optimal value of $\rm{N_{skip}}=650$ is adopted, as a good compromise between charge resolution and pixel exposure to dark current.  A 10$\times$10 pixel binning\,\footnote{Pixel binning is an operating mode of the CCD where the charge of several pixels is summed before being read out. An n$\times$m binning corresponds to summing the charge of n pixels in the horizontal direction and m pixels in the vertical direction.} is used for the readout. By binning, charge from a point-like energy deposit distributed by diffusion over several physical pixels is summed before measurement, improving the signal-to-noise ratio. The binning size is optimized using the measured value of $\mathrm{\sigma_{xy}}$ (see\,\cite{PhysRevD.105.062003} for the measurement method) so that DM interactions are most likely contained in a single binned pixel. For the remainder of this text, the term pixel is used to describe a 10$\times$10 bin of pixels (i.e., 150$\times$150 $\mu \rm{m^2}$). A continuous readout mode, where images of 640$\times$840 ($\rm{N_{col}\times N_{row}}$) pixels are taken subsequently, each beginning immediately after the end of the previous one, is implemented for Science Run 1 (SR1), resulting in the same exposure time for each pixel. 
In Science Run 2 (SR2), only a fraction of the CCD is read out (640$\times$110 pixels), and the charge in the CCD is cleared between consecutive images. In this mode the pixel's exposure time increases linearly as a function of row, and a lower average charge accumulates during the pixel exposure ($\approx$0.0033\,e$^-$/pixel/image in SR2  vs $\approx$0.012\,e$^-$/pixel/image in SR1) resulting in a lower rate of pixels with charge $>$ 1~e$^-$. As the LBC is still in its commissioning phase, this level of dark current ($\approx$20\,e$^-$/mm$^2$/day) is several times higher than the lowest reached in CCDs\,\cite{Damic_dc_2019,sensei2020}, but sufficiently low to perform a sensitive search for DM. 

The following procedure is used to reduce and calibrate the raw CCD images. First, the pixel charge is obtained by averaging the NDCMs. Then, a DC offset, or pedestal, introduced by the electronics chain is subtracted. The pedestal value is determined row-by-row from a Gaussian fit of the charge distribution's most prominent peak, comprised of pixels with zero charge. 

\begin{figure}[ht!]
    \centering
    \includegraphics[width=\linewidth]{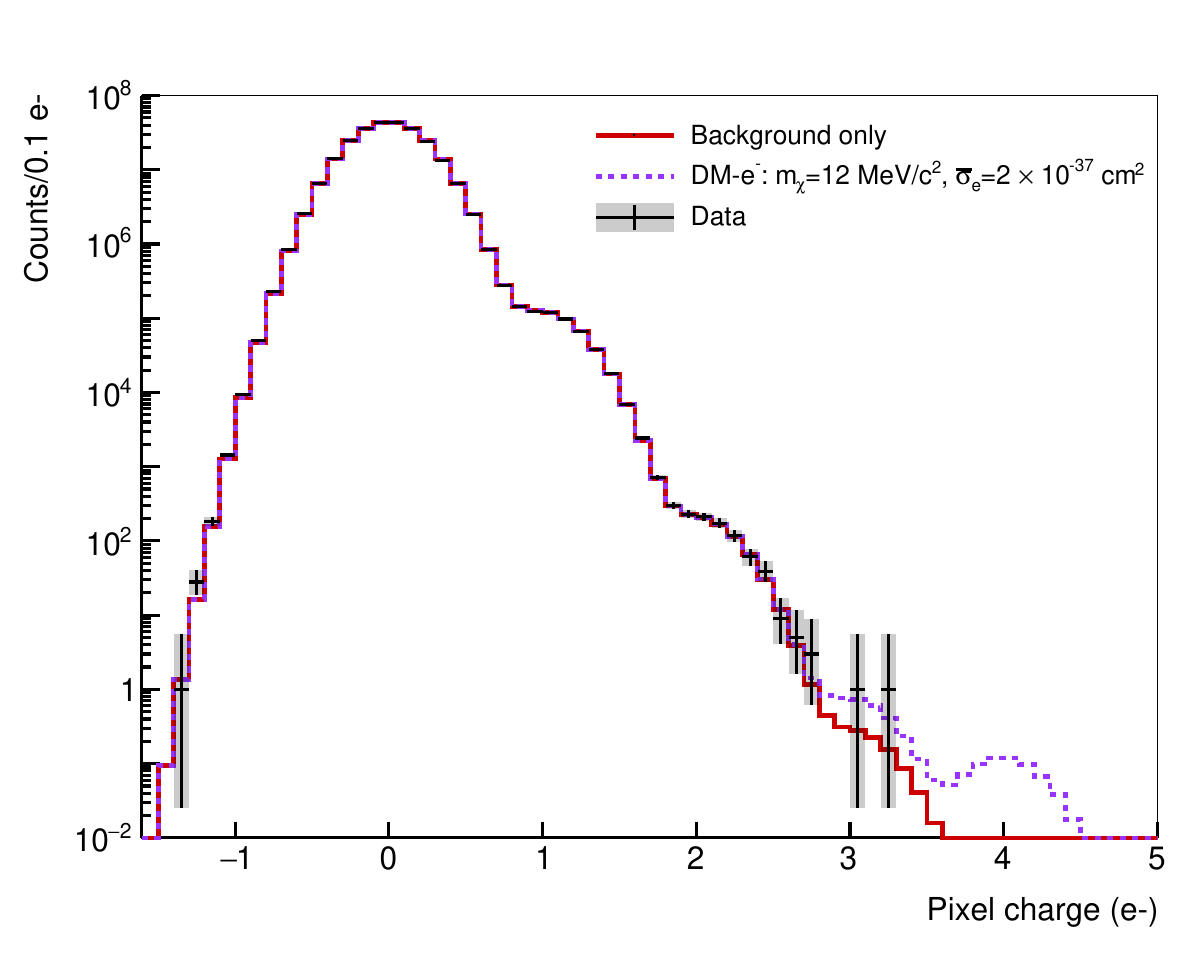}
    \caption{Example of a pixel charge distribution with peaks corresponding to individual charges. This distribution, with pixels from the U amplifier of the SR2 data set, constitutes about 40\% of the full data set used for the DM search. The red line is the fit result for the background-only hypothesis (no DM-e$^-$). The dashed violet line is the expectation for background plus a DM-e$^-$ heavy-mediator model with $m_\chi$=12\,MeV$/c^2$ and \sige\ = $2 \times 10^{-37}~\rm{cm^2}$, which is equal to the 90\% CL limit value obtained at this mass. }
    \label{fig:PCD}
\end{figure}

The calibration constant, which converts the measured analog-to-digital units (ADU) into the number of electrons\,\footnote{For the sake of simplicity, we use the term electrons to indicate charge carriers detected in the CCD. However, holes are held in the pixels of the p-channel CCD used for this measurement.}, is obtained by fitting a Gaussian function convolved with a Poisson distribution\,\cite{Compton-DAMICM}. The charge resolution, $\sigma_{\text{res}}\approx$ 0.2~\,$e^-$, is estimated from the standard deviation of the Gaussian fit. The U and L amplifiers in each CCD are calibrated independently.  An example of a calibrated pixel charge distribution is shown in Fig.~\,\ref{fig:PCD} where the peaks correspond to 0, 1 and 2\,e$^-$, from left to right.  
We then identify energy deposits, which may extend over more than one pixel. Adjacent pixels with charge $\ge3\,\sigma_{\text{res}}$ are grouped together as a cluster if at least one pixel has $\ge2$\,e$^-$. Clusters or single isolated pixels with charge $>7$\,e$^-$ are excluded from further analysis since the probability that they originate from a DM interaction is negligible for the DM mass range of interest. We also exclude the 10 trailing pixels in the horizontal and vertical directions to account for charge transfer inefficiencies. Monte Carlo simulations show that the efficiency for a dark matter signal with charge $\le 7$\,e$^-$ is not affected by this procedure. The clustering selection rejects about $6\times 10^{-5}$ of the pixels. 
Defects in the CCD may release charge during the readout process, appearing as ``hot'' pixels and columns~\cite{2001sccd.book.....J}. 
To identify these defects, we parametrize the 1e$^-$ rate as a function of column number $i_{\rm{col}}$ with a second-order polynomial $P_{pol}(i_{\rm{col}})$ and then tag columns with a rate exceeding the parametrization by more than 2~$\sigma$. 
We also use a dedicated data set of 13 images with 3-hour exposures to identify high-charge pixels recurring in multiple images. Columns corresponding to the identified defects are then excluded from the analysis. These criteria select 80.4\% of the pixels with an efficiency which does not depend on the pixel charge. 
Finally, we identify the location of artifacts in the serial register as columns with a sizable reduction in dark current, {\it i.e.} with 1e$^-$ rate $>2~\sigma$ below the parametrization $P_{pol}(i_{\rm{col}})$. 
While moving through a serial register, charge transfer may be delayed by the presence of a trap or a local anomaly in the electric fields, effectively changing the expected pixel charge distribution from a DM interaction. Thus, we select only portions of the CCD active area not affected by serial register artifacts.
Several such artifacts are identified in one of the two prototype CCDs, which is therefore excluded from further analysis. For the remaining CCD, pixels in the L side with $i_{\rm{col}} ~ >$ 74 are rejected.   
After applying the selection criteria 3.68$\times10^8$ pixels remain, corresponding to a final integrated exposure for the DM search of 85.23\,g-days (45.26\,g-days for SR1 and 39.97\,g-days for SR2). No pixel with charge $\ge4$\,e$^-$ and $\le 7$\,e$^-$ is present in this data set\footnote{In a preliminary analysis of the SR1 dataset~\cite{damic_idm2022} one $4$\,e$^-$ event was found. This event is removed from the present analysis by the rejection criterion for serial register artifacts.}, improving by one order of magnitude previous limits in silicon at these charge multiplicities\,\cite{sensei2020}. 

To place an upper limit on the DM signal a joint binned-likelihood fit is performed on four pixel distributions (one for each amplifier in each of the two science runs).
An entry in these distributions corresponds to the value of one unmasked pixel (out of $\rm{N_{pix}}$) in one image (out of $\rm{N_{im}}$).
The charge of a pixel may come from different background sources. Radiogenic backgrounds modeled in Ref.~\cite{PhysRevD.105.062003} may contribute at most 0.02 pixels at each charge multiplicity between 1 and $7$\,e$^-$, and is thus negligible. Other backgrounds~\cite{2001sccd.book.....J} include thermally-generated or stress-induced dark counts, spurious clock-induced charge, and photoabsorption of light from the readout amplifiers.
These are all Poisson processes that act as a source of uncorrelated, single electrons with mean value of $\lambda_i$ for the $i$th pixel in the CCD.
Since different pixels can have different exposures, and most background sources depend on the location of the pixel in the CCD (e.g., local stress, physical distance from the readout amplifiers, number of charge transfers to the readout amplifier, etc.), $\lambda_i$ varies for each pixel.
To estimate $\lambda_i$ for a given pixel $i$, we perform a fit to the 0 and 1 electron peaks in the distribution of the $\rm{N_{im}}$ charge values of pixel $i$. We then build the background-only hypothesis $B$ by adding the contribution from every pixel in the data set:

\begin{equation}
    \label{eq:bonly}
    B(p|\lambda_i,\sigma_\text{res}) = \sum_{i=0}^{\mathrm{N_{pix}}} \mathrm{N_{im}} \sum_{n_q=0}^{\infty} \text{Pois}(n_q|\lambda_i)\text{Gaus}(p|n_q,\sigma_\text{res}),
\end{equation}

\noindent where $p$ is the observed charge value given $n_q$ electrons collected by the pixel in an image. $\text{Pois}(n_q|\lambda_i)$, the Poisson probability of obtaining $n_q$ given $\lambda_i$, is the amplitude of Gaussian functions $\text{Gaus}(p|n_q,\sigma_{\rm res})$ with mean $n_q$ and standard deviation $\sigma_{\rm res}$ to model the readout noise. 

A DM flux of particles with $m_\chi$ and $\bar\sigma_e$ may contribute $j$ charges in a pixel with exposure $\epsilon_i$ with probability distribution $S(j|m_\chi , \bar\sigma_e, \epsilon_i)$. The fit function $F$ which includes both the signal and background model is then given by:

\begin{widetext}
\begin{equation}
    \label{eq:cumulPCD}
    F(p|m_\chi,\bar\sigma_e,\epsilon_i,\lambda_i,\sigma_\text{res}) =  \sum_{i=0}^{\mathrm{N_{pix}}}
    \mathrm{N_\mathrm{im}} \sum_{n_q=0}^{\infty} \left [ \sum_{j=0}^{n_q}  S(j|m_\chi,\bar\sigma_e, \epsilon_i) \text{Pois}(n_q-j|\lambda_i-\lambda_{S, i}) \right ] \text{Gaus}(p|n_q,\sigma_\text{res}).
\end{equation}
\end{widetext}

A DM signal that contributes to the one-electron counts would make our empirical procedure for the background model overestimate $\lambda_i$. We correct for this effect by subtracting from $\lambda_i$ in Eq.~\ref{eq:cumulPCD} the number of one-electron counts contributed by a given signal $S$ in the ith pixel, $\lambda_{S, i}$. Note that for $m_\chi <$1\,MeV$/c^2$, where interactions produce at most one electron, the signal is indistinguishable from the background model and only an upper limit on the interaction rate can be placed.

\begin{figure*}[t]%[h!]
    \centering
	\hspace{-0.2cm}
        \includegraphics[width=0.45\linewidth]{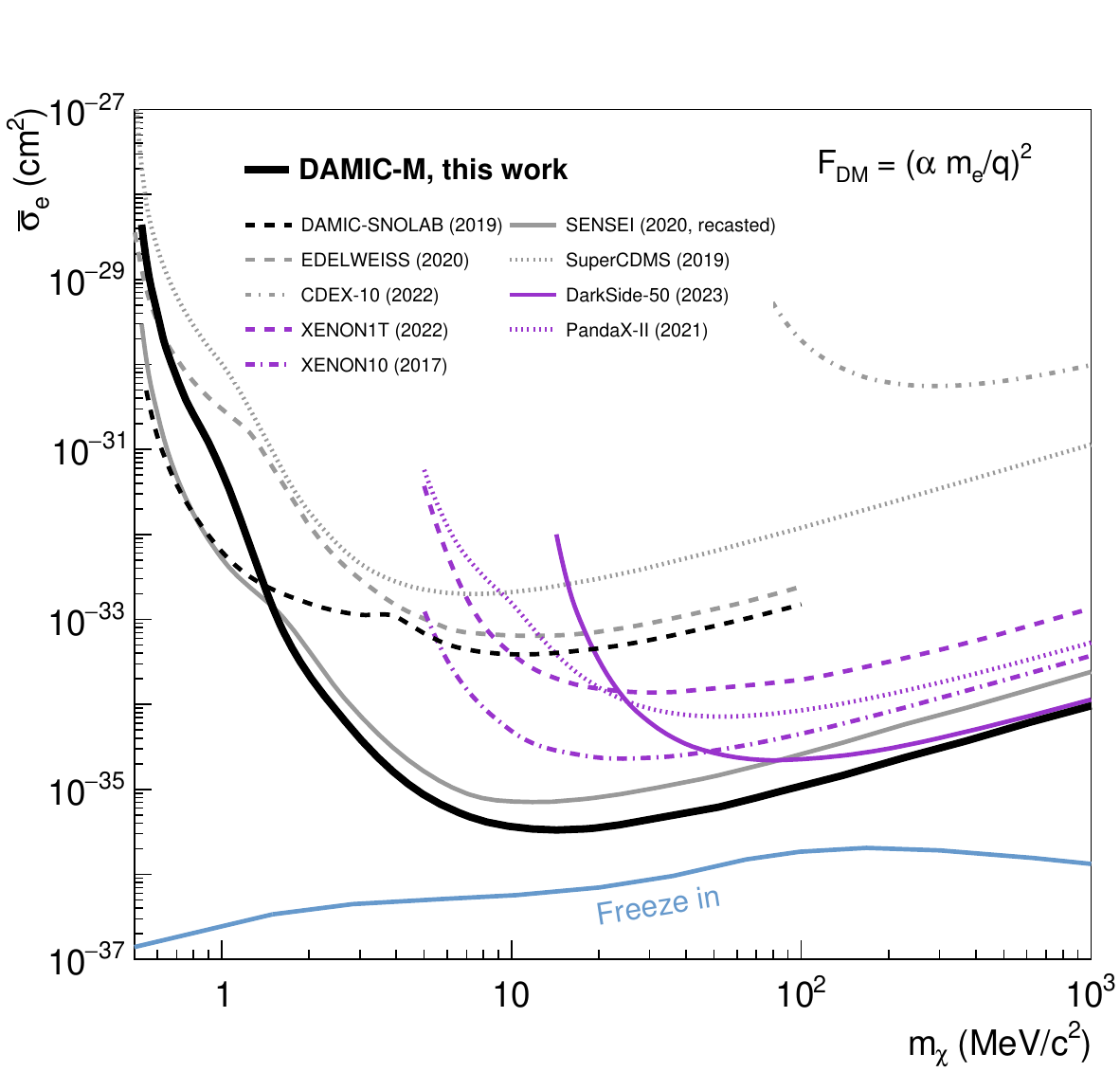}
        \hspace{0.6cm}
   	\includegraphics[width=0.45\linewidth]{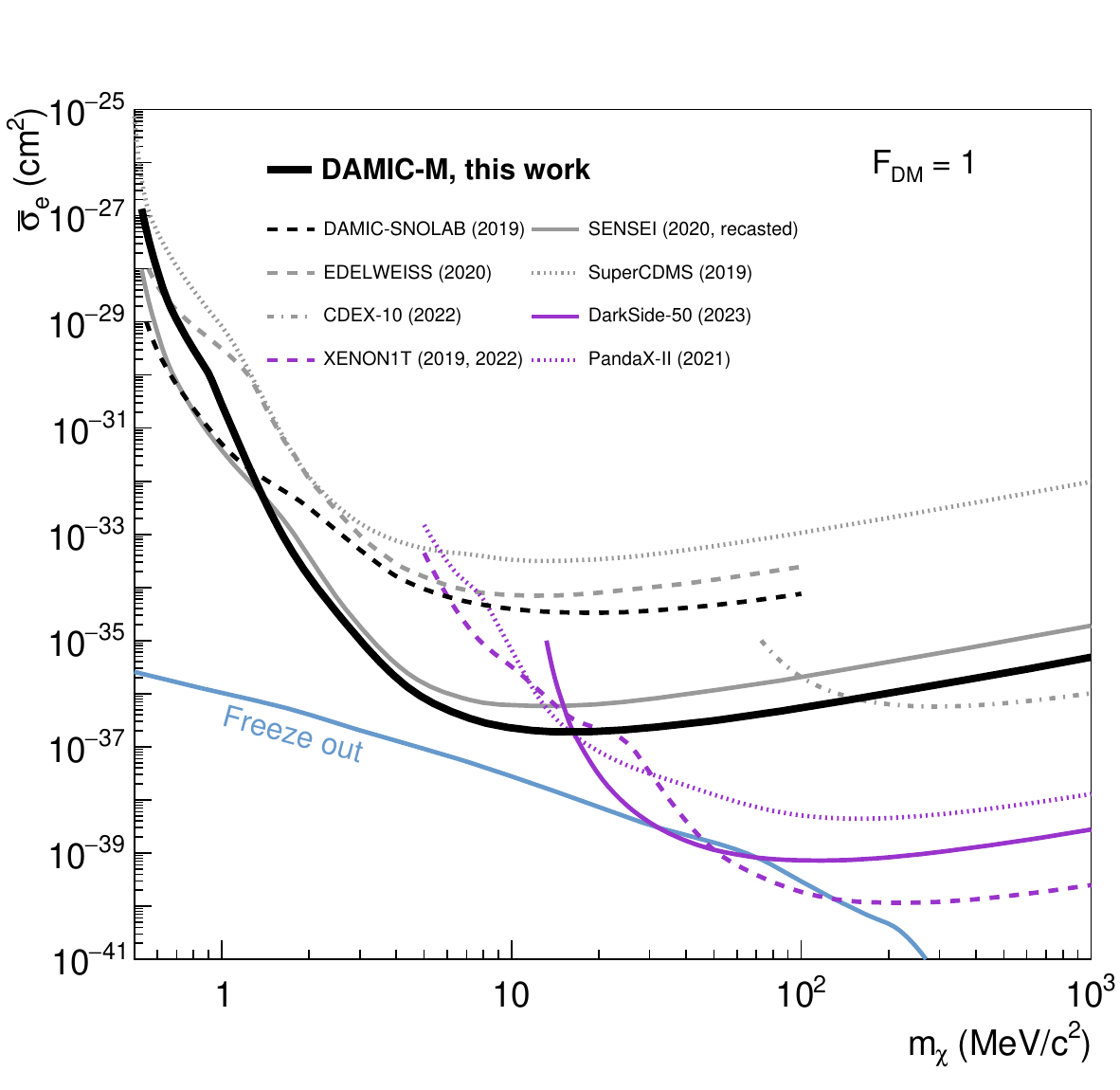}
    \caption{DAMIC-M 90\% C.L. upper limits (solid black) on DM-electron interactions through a ultra-light mediator (left) and heavy mediator (right). Also shown are current best direct-detection limits from other experiments, DAMIC-SNOLAB\,\cite{DAMIC:2019dcn} (dashed black), SENSEI\,\cite{sensei2020} (solid gray), EDELWEISS\,\cite{edelweiss2020} (dashed gray), SuperCDMS\,\cite{superCDMS2018,*PhysRevLett.122.069901} (dotted gray), CDEX-10\, \cite{cdex10} (dot-dashed gray), DarkSide-50\,\cite{PhysRevLett.130.101002} (solid violet), XENON1T combined result from \cite{xenon1t2019, xenon1t-SE-2022} (dashed violet), PandaX-II\,\cite{PandaX-2021} (dotted violet), and a limit obtained from XENON10 data in Ref.\,\cite{xenon10_2017} (dash-dotted violet). Theoretical expectations assuming a DM relic abundance from freeze-in and freeze-out mechanisms are also shown in light blue\,\cite{USCosmicVisions2017}.}
\label{fig:upperlimits}
 \end{figure*}
 
The DM signal $S$ is computed with Eq.\,\ref{eq:semicondrate} using \texttt{QEdark}\,\cite{QEDark} for $f_c$ and a standard Maxwell-Boltzmann velocity distribution for the $\eta$ factor with parameters of the DM density profile in the galactic halo as recommended in Ref.\,\cite{PhystatDM}\footnote{Local dark matter density $\rho_\chi=0.3~$GeV/c$^2$/cm$^3$, mean $v_0=238.0$\,km/s, escape velocity $v_{\text{esc}}=544$\,km/s, and mean periodic Earth velocity $v_{\text{E}}=253.7$\,km/s}. We obtain the DM interaction rate as a function of discrete ionization charges starting with the semi-empirical electron-hole pair creation probabilities $P_{pair} (n_q|E_e)$ from the charge yield model of Ref.~\cite{PhysRevD.102.063026}. 
A Monte Carlo (MC) simulation is then used to include the detector response. Charge is injected uniformly across the sensitive volume of the CCD and diffused on the pixel array with a Gaussian transverse variance $\mathrm{\sigma^2_{xy}(z)}=-a\ln|1-b\mathrm{z}|\cdot(\alpha + \beta E_e)^2$, with parameters $a = 2108~\mu\rm{m}^2$, $b = 1.98 \times 10^{-4}~ \mu\rm{m^{-1}}$, $\alpha=0.859$, and $\beta= 0.0067~\rm{keV^{-1}}$ calibrated with cosmic rays in a surface lab (see e.g. Ref.\,\cite{PhysRevD.105.062003}). A 10x10 binning of the simulated CCD array is then performed to match the data-taking conditions. This procedure, repeated for different DM masses, yields the signal $S$ in Eq.\,\ref{eq:cumulPCD}.

We then fit the model of Eq.\,\ref{eq:cumulPCD} to data by maximizing a binned log-likelihood $\mathcal{L}$, which assumes Poisson bin content. 
Since the charge resolution is set by the individual amplifier's readout noise, $\sigma_\text{res}$ is an independent free parameter for each of the four pixel distributions.
An example of a fit result for the background-only hypothesis (\sige\ =0) is shown in Fig.\,\ref{fig:PCD} for the U amplifier of the SR2 data set. Also shown for illustration is the expected distribution for the background model plus a DM heavy-mediator signal with $m_\chi$=12\,MeV$/c^2$ and \sige\ = $2 \times 10^{-37}~\rm{cm^2}$ equal to the 90\% CL limit value obtained at this mass.

No preference is found for a DM signal and exclusion limits are derived accordingly. 
We use the approach of Ref.\,\cite{asymptoticlikelihood} and the profile likelihood ratio test statistic, $t_\sigma = -2\log\lambda(\sigma)$ where $\lambda(\sigma)$ is the profile likelihood ratio, at each DM mass. The DAMIC-M $90\%~$C.L. exclusion limits for heavy\,(right) and ultra-light\,(left) mediator sub-GeV DM are shown in Fig.\,\ref{fig:upperlimits}. We find these limits to be within the expected 68\% sensitivity band as estimated by MC simulations.
Results from other direct detection experiments are also shown in Fig.\,\ref{fig:upperlimits}, where the limit from SENSEI\,\cite{sensei2020}, which also uses skipper CCDs, was recasted for proper comparison by using the same halo parameters\,\cite{PhystatDM} and charge yield model\,\cite{PhysRevD.102.063026} adopted in this analysis. Theoretical expectations for models which reproduce the correct DM relic abundance by thermal ``freeze-out" of DM annihilation into standard model particles (heavy mediator) during the early universe or ``freeze-in" of standard model particles annihilation into DM (ultra-light mediator)~\cite{QEDark} are also shown in Fig.\,\ref{fig:upperlimits}.

Several cross-checks of the analysis procedures have been performed. We verify with dedicated data sets that pixel charge multiplicities relevant to this analysis are not altered by charge transfer inefficiency.
A more elaborate 2-D analysis of the pixel charge distribution, which slightly improves charge resolution by exploiting noise correlation between symmetric pixels on the U and L side, is employed. Independent cross-checks have been performed at every step in the analysis, starting from the low-level image processing to the generation of the data pixel distribution, the identification of defects, the modeling of the DM signal, and the extraction of the DM signal upper limit. Consistent results are obtained in all of these checks, indicating no major systematic effect in our procedure.  
We evaluate theoretical uncertainties associated with the calculation of the DM-e$^-$ interaction rate by using \texttt{DarkELF}\,\cite{Knapen_2022} and \texttt{EXCEED-DM}\,\cite{Griffin_2021, exceedDMpublished} predictions for the signal $S$. The corresponding limits are in general worse than the \texttt{QEdark}-based results of Fig.~\ref{fig:upperlimits}, up to a factor of 60 at low DM masses. 
%The corresponding limits are between 2 and 60 times worse than the \texttt{QEdark}-based results of Fig.~\ref{fig:upperlimits}, with the largest difference at low DM masses. 
Thus approximations in the theoretical models (e.g. no in-medium screening effects in \texttt{QEdark}) have significant impact. We use \texttt{QEdark} as the reference theoretical model for proper comparison with previous and forthcoming results from other experiments and include in the Supplemental Material the limits derived with the other models.

This DAMIC-M search for DM particles of mass between 0.53 and 1000\,MeV$/c^2$ excludes unexplored regions of parameter space in mass ranges [1.6,1000]\,MeV$/c^2$ for an ultra-light mediator and [1.5,15.1]\,MeV$/c^2$ for a heavy mediator. 

Efforts are ongoing to significantly decrease the dark current in upcoming upgrades to the LBC, including the deployment of the final DAMIC-M CCD modules with lower mechanical stress, better shielding from infrared radiation, and readout electronics with lower noise.

\begin{acknowledgments}
We thank the LSM for their support in the installation and operation of the detector underground.
The DAMIC-M project has received funding from the European Research Council (ERC) under the European Union's Horizon 2020 research and innovation programme Grant Agreement No. 788137, and from NSF through Grant No. NSF PHY-1812654. 
The work at University of Chicago and University of Washington was supported through Grant No. NSF PHY-2110585. This work was supported by the Kavli Institute for Cosmological Physics at the University of Chicago through an endowment from the Kavli Foundation. 
We thank the College of Arts and Sciences at University of Washington for 
contributing the first CCDs to the DAMIC-M project.
IFCA was supported by project PID2019-109829GB-I00 funded by MCIN/ AEI /10.13039/501100011033.
The Centro At\'{o}mico Bariloche group is supported by ANPCyT grant PICT-2018-03069.
The University of Z\"{u}rich was supported by the Swiss National Science Foundation.
The CCD development work at Lawrence Berkeley National Laboratory MicroSystems Lab was supported in part by the Director, Office of Science, of the U.S. Department of Energy under Contract No. DE-AC02-05CH11231. We thank Teledyne DALSA Semiconductor for CCD fabrication. We thank Tanner Trickle for useful discussions on the use of \texttt{EXCEED-DM}.
\end{acknowledgments}

% references
\bibliographystyle{style/h-physrev}
\bibliography{apssamp}% Produces the bibliography via BibTeX.

\end{document}

\maketitle

\begin{figure*}[h!]%[h!]
    \centering
	\hspace{-0.2cm}
        \includegraphics[width=0.45\linewidth]{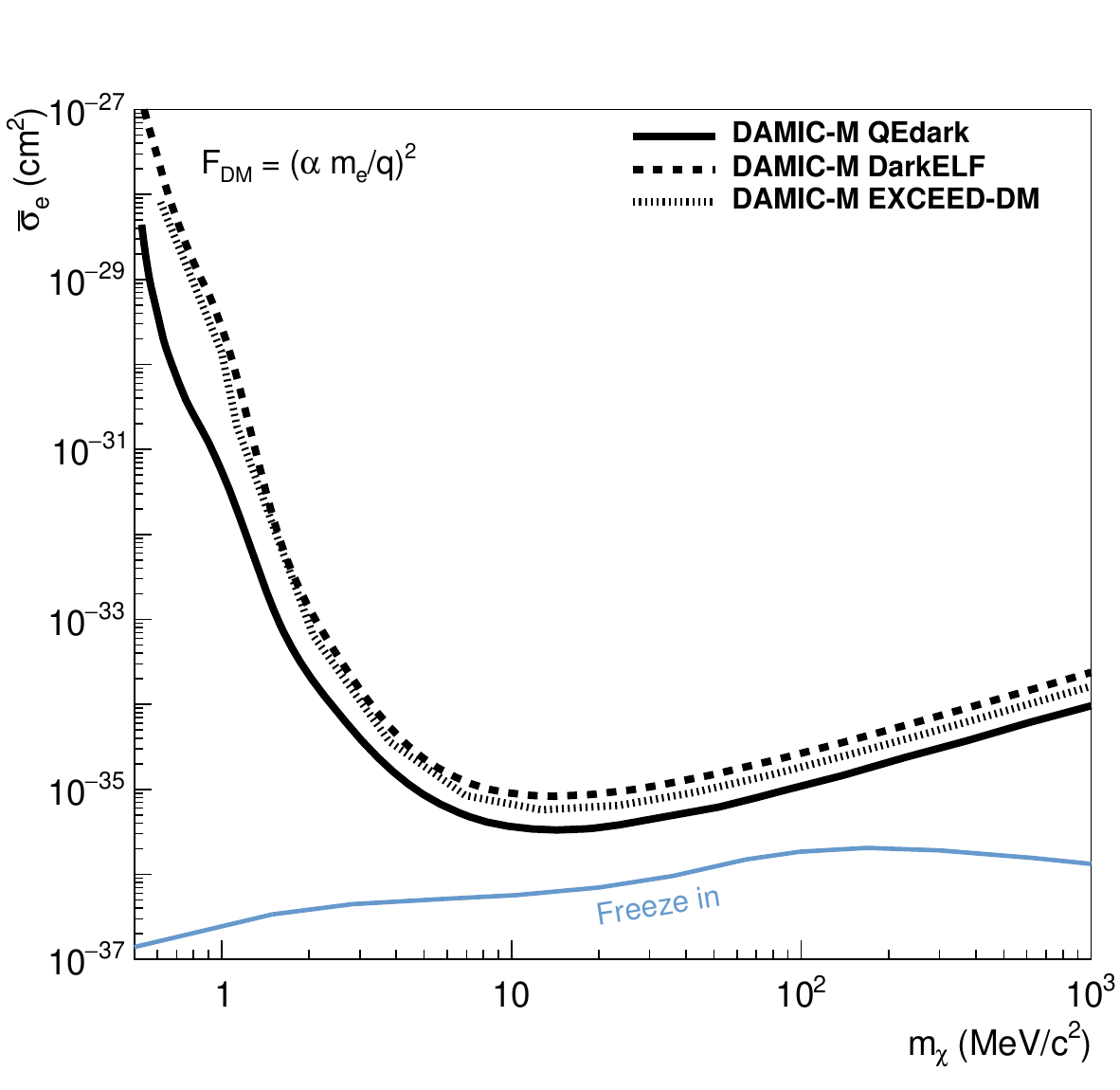}
        \hspace{0.6cm}
   	\includegraphics[width=0.45\linewidth]{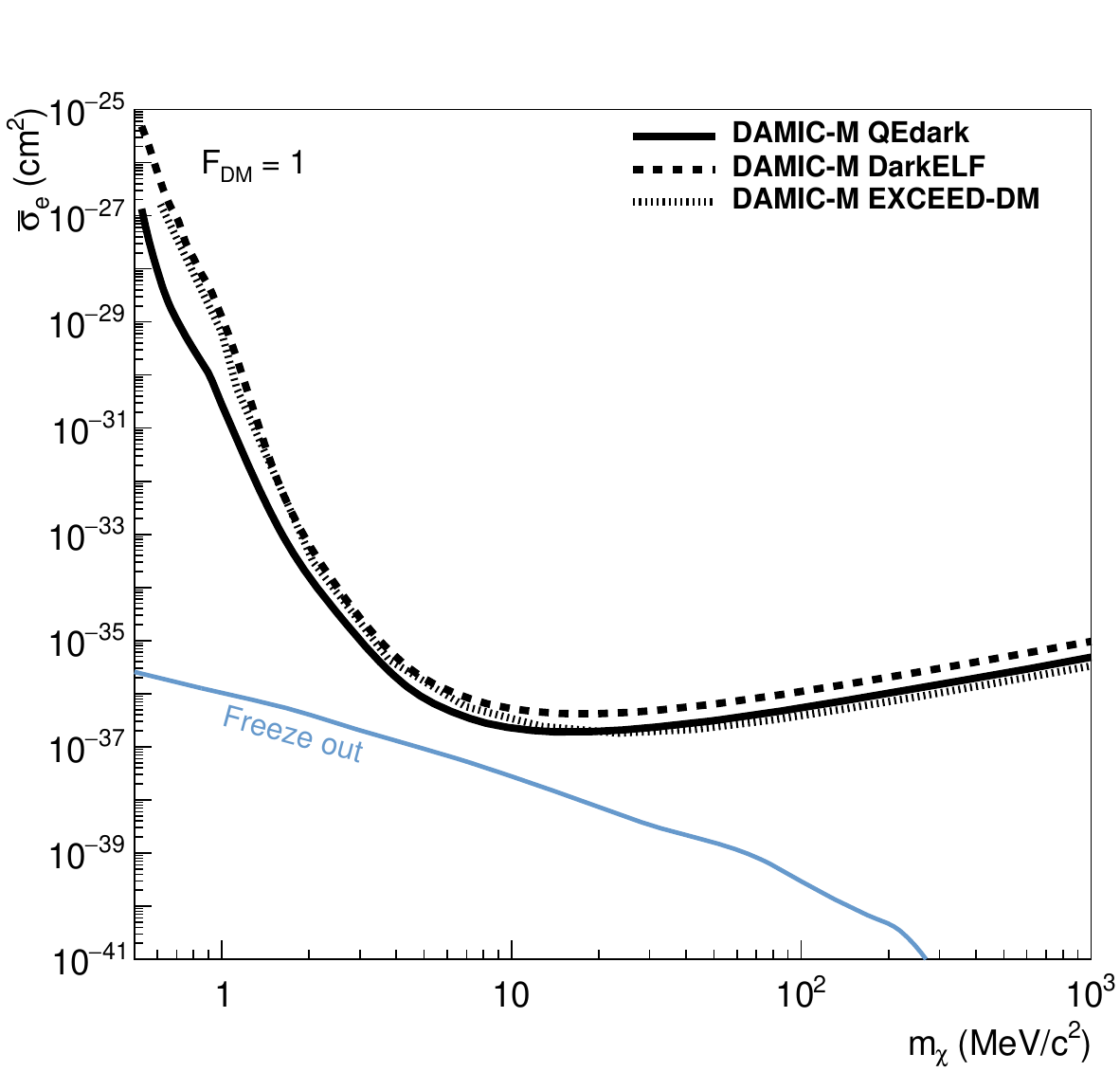}
    \caption{DAMIC-M 90\% C.L. upper limits on DM-electron interactions through a
ultra-light mediator (left) and heavy mediator (right) obtained with the \texttt{QEdark} [21] (solid), \texttt{DarkELF} [24] (dashed), and \texttt{EXCEED-DM} [23,47] (dotted) theoretical
models. The GPAW method is used for the \texttt{DarkELF} calculation. The recommended Si electronic  configuration file (10.5281/zenodo.7246141) and numerically computed dielectric function are used for \texttt{EXCEED-DM}. }
\label{fig:upperlimits}
 \end{figure*}